\documentclass[DLO3.tex]{article}

\usepackage{arxiv}

\usepackage[utf8]{inputenc} 
\usepackage[T1]{fontenc}    
\usepackage{hyperref}       
\usepackage{url}            
\usepackage{booktabs}       
\usepackage{amsfonts}       
\usepackage{nicefrac}       
\usepackage{microtype}      
\usepackage{lipsum}
\usepackage{graphicx}
\usepackage{tabularx,ragged2e,booktabs,caption}
\usepackage{graphicx} 
\usepackage{graphics}
\usepackage{subfigure}
\usepackage[title]{appendix}
\usepackage{xcolor}

\title{Recurrent U-net: Deep learning to predict daily summertime ozone in the United States}

\author{
	Tai-Long He \\
	Department of Physics\\
	University of Toronto\\
	\texttt{the@physics.utoronto.ca} \\
	\And
	Dylan B. A. Jones \\
	Department of Physics\\
	University of Toronto\\
	\texttt{dbj@atmosp.physics.utoronto.ca} \\
	\And
	Binxuan Huang \\
	School of Computer Science  \\
	Carnegie Mellon University \\
	\texttt{binxuanh@cs.cmu.edu} \\
	\And
	Yuyang Liu \\
	Department of Computer Science  \\
	University of Toronto \\
	\texttt{yuyang@cs.toronto.edu} \\
	\And
	Kazuyuki Miyazaki \\
	Jet Propulsion Laboratory \\
	California Institute of Technology \\
	\texttt{kazuyuki.miyazaki@jpl.nasa.gov} \\
	\And
	Zhe Jiang \\
	School of Earth and Space Sciences \\
	University of Science and Technology of China \\
	\texttt{zhejiang@ustc.edu.cn} \\
	\And
	E. Charlie White \\
	Department of Physics \\
	University of Toronto \\
	\texttt{cwhite@physics.utoronto.ca} \\
	\And
	Helen M. Worden \\
	Atmospheric Chemistry Observations and Modelling \\
	National Center for Atmospheric Research \\
	\texttt{hmw@ucar.edu} \\
	\And
	John R. Worden \\
	Jet Propulsion Laboratory \\
	California Institute of Technology \\
	\texttt{john.r.worden@jpl.nasa.gov}  \\
}

\begin{document}
	\maketitle
	
	\begin{abstract}
		
		We use a hybrid deep learning model to predict June-July-August (JJA) daily maximum 8-h average (MDA8) surface ozone concentrations in the United States. A set of meteorological fields from the ERA-Interim reanalysis as well as monthly mean NOx emissions from the Community Emissions Data System (CEDS) inventory are selected as predictors. Ozone measurements from the US Environmental Protection Agency (EPA) Air Quality System (AQS) from 1980 to 2009 are used to train the model, whereas data from 2010 to 2014 are used to evaluate the performance of the model. The model captures well daily, seasonal and interannual variability in MDA8 ozone across the United States, with $r^2 = 0.83$. The model demonstrates high predictive skill in the eastern United States and on the West Coast, but low skill in the Intermountain West, where $r^2 \approx 0.4$. The mean error on the daily MDA8 ozone is $-1.14 \pm 1.94$ ppb for the United States. Feature maps show that the model captures teleconnections between MDA8 ozone and the meteorological fields, which are responsible for driving the MDA8 ozone dynamics. We used the model to evaluate recent trends in NOx emissions in the United States and found that the trend in the EPA emission inventory produced the largest negative bias in MDA8 ozone between 2010-2016. The top-down emission trends from the Tropospheric Chemistry Reanalysis (TCR-2), which is based on satellite observations, produced MDA8 ozone predictions in best agreement with observations across the United States. However, in urban regions (defined by high emissions of NOx), the trend in AQS NO$_2$ observations provided ozone predictions in agreement with observations, whereas in rural regions the top-down satellite-derived trends produced the best agreement with the observations. In both rural and urban regions the EPA trend resulted in the largest negative bias in predicted ozone. Our results suggest that the EPA inventory is overestimating the reductions in NOx emissions and that the satellite-derived trend reflects the influence of reductions in anthropogenic NOx emissions as well as changes in background NOx. Our results demonstrate the significantly greater predictive capability that the deep learning model provides over conventional atmospheric chemical transport models for air quality analyses.

	\end{abstract}


	\section{Introduction}
	\label{sec:intro}
	Tropospheric ozone is a major air pollutant and a greenhouse gas. It is produced photochemically by the oxidation of hydrocarbons in the presence of nitrogen oxides (NOx = NO + NO$_2$). As a consequence, ozone abundances near the surface are at a maximum in summer. Due to its high oxidative capability, high abundances of ozone near the surface are associated with adverse impacts on human health and crop yield. There are significant variations in surface ozone in the United States on both short and long time scales reflecting the influence of meteorology, non-linearity in the ozone chemistry, and changes in the emissions of ozone precursor gases. Atmospheric models used to simulate the distribution of ozone typically do not reproduce the observed long-term trend in tropospheric ozone. Furthermore, these models tend to overestimate summertime surface ozone abundances in the United States. For example, in an evaluation of 16 global models and one hemispheric model, Reidmiller et al.~\cite{reidmiller} found that the models overestimated summertime maximum daily 8-hour average (MDA8) ozone ozone in the eastern United States by 10--20 ppb. Recently, Travis et al.~\cite{travis2019} found that the high-resolution regional version of the GEOS-Chem model has a MDA8 bias of +5 ppb, which is reduced to +1 ppb when the model evaluation is restricted to afternoon hours under dry conditions, with proper accounting for the vertical gradient of ozone in the lowest model layer. The Travis et al.~\cite{travis2019} results highlight the challenges of using conventional chemical transport models for air quality applications. Here, we apply a hybrid deep learning model to predict June-July-August (JJA) MDA8 ozone in the United States using meteorological and chemical predictors. Compared to existing atmospheric models, the deep learning approach offers superior predictive capability for summertime ozone, better accounting for the coupling between meteorology and emissions \cite{fiore-b}. 
	
	Stagnant weather conditions and high surface temperatures are favorable for the occurrence of extreme surface ozone episodes in summer. Previous studies have used statistical methods to investigate the relationship between large-scale atmospheric circulation patterns and summertime surface ozone \cite{gardner, lshen}. Recent achievements in deep learning \cite{deeplearning} over the past few years show that empirical models are able to learn both spatial and temporal patterns in the input data. The application of deep learning models has achieved great success in computer vision and natural language processing. As discussed in \cite{reichstein}, deep learning approaches have the potential to improve our predictive ability and understanding in a wide range of challenges we have in Earth science. There have been several previous studies using deep learning models for classification problems in the Earth Sciences \cite{lee1990, benediktsson, gomez, campsvalls}. For example, Liu et al. \cite{liu2016} used a deep convolutional neural network to detect extreme events in a climate dataset. However, not many state-of-the-art deep learning architectures have been employed for predictive applications in the Earth Sciences. In this study, we propose a deep learning model that is able to learn the spatiotemporal dynamics of summertime ozone in the United States. We show that the model captures well both long-term and short-term variability in MDA8 ozone over the United States, and that the model is able to provide predictions where no \textit{in situ} observations are available. We also utilize the fitted model to conduct a qualitative analysis of the teleconnections between MDA8 and its predictors. 
	
	Atmospheric NOx is a key precursor of tropospheric ozone. Due to air pollution regulations, NOx emissions have declined significantly since the 1990s. However, there is uncertainty about the recent trends in NOx emissions in the United States. NOx emission estimates inferred from satellite observations (referred to as top-down estimates) suggest that there has been a slowdown in the reduction rate since 2009, compared to the bottom-up emission inventory reported by the US Environmental Protection Agency (EPA) \cite{jiang}. However, it has also been suggested \cite{silvern} that the slowdown in the reduction rate in the satellite-derived emission estimates does not indicate a discrepancy with the EPA emission inventory, but instead is due to the increasing relative influence of non-anthropogenic NOx emissions on atmospheric NOx as captured by the satellite measurements. It was also suggested in \cite{li_2019_trends} that the satellite-derived trends are consistent with the trends in surface observations of NOx in high emission regions and that the discrepancy between the top-down and bottom-up trends are due to non-linearity in the relationship between NOx emissions and the satellite observations of NO$_2$ in rural regions. Here we use the deep learning model to evaluate the recent trends in NOx emissions in the United States. The deep learning model is independent of the chemical errors that are typically found in atmospheric chemical transport models used in this type of evaluation (e.g. \cite{silvern}). We demonstrate that the model is therefore an ideal tool for air quality assessment studies.

	\section{Summertime ozone predictors}
	\label{sec:predictors}
	
	Large-scale atmospheric circulation patterns, sea surface temperatures (SSTs), and sea level pressure (SLP) have an impact on year-to-year variability of summertime ozone in the eastern United States \cite{lshen}. The dynamical patterns in these meteorological fields in the previous spring have been shown to have relatively high correlation with summertime MDA8 in the eastern United States, and could be used as predictors in statistical models. Reductions in NOx emissions have driven the negative trend in surface ozone in the United States \cite{cooper}. Simulations suggest that the 95th percentile of summertime ozone would have slightly increased over this period (1990--2010) in absence of NOx emission regulations \cite{lin-c}. We have therefore selected the following MDA8 ozone predictors, focusing on the June-July-August (JJA) period: anthropogenic emissions of NOx and non-methane volatile organic compounds (NMVOCs), mean sea level pressure (MSLP), geopotential at 500 hPa level (Z), downward shortwave radiation (SSRD), sea surface temperature (SST), 2-meter temperature (T2M), and 2-meter dew point (D2M). The NOx emissions are separated into seven emissions sectors that agriculture (AGR), the power industry (ENE), the manufacturing industry (IND), residential and commercial (RCO), international shipping (SHP), surface transportation (TRA), and waste disposal (WST). The sector-based NOx emissions provide geospatial information to the neural networks, which helps with the regression and localization of ozone levels.

	\section{Data}
	\label{sec:data}
	The meteorological data are from the ERA-Interim reanalysis \cite{dee} from the European Centre for Medium-Range Weather Forecasts (ECMWF), which have been regridded to a horizontal resolution of $1.5^{\circ} \times 1.5^{\circ}$. The NOx emissions are from the Community Emissions Data System (CEDS) inventory \cite{ceds}. All the data are cropped to a regional domain extending from 0$^{\circ}$N to 72$^{\circ}$N, and from 180$^{\circ}$W to $0^{\circ}$E to encompass the North Pacific and the North Atlantic, where strong linkages were found between ocean forcing and summertime climate in the eastern United States \cite{lshen, sutton1, sutton2, gill}.
	
	MDA8 ozone was estimated from ozone measurements from the EPA Air Quality System (AQS) (https://www.epa.gov/aqs). The MDA8 ozone were aggregated to $3^{\circ} \times 3^{\circ}$ grid boxes. With a 37-year AQS MDA8 record extending from 1980 to 2016, we used JJA data from 1980 to 2009 as the training data set. During the training process, the last 15\% of the data were used as a validation data set, in order to prevent overfitting the model. The MDA8 data from 2010 to 2014 were used as the testing set to evaluate the performance of the model. 
	
	\section{A deep learning model to predict MDA8}
	\label{sec:model}
	
\begin{figure}
		\centering
		\includegraphics[width=.60\textheight]{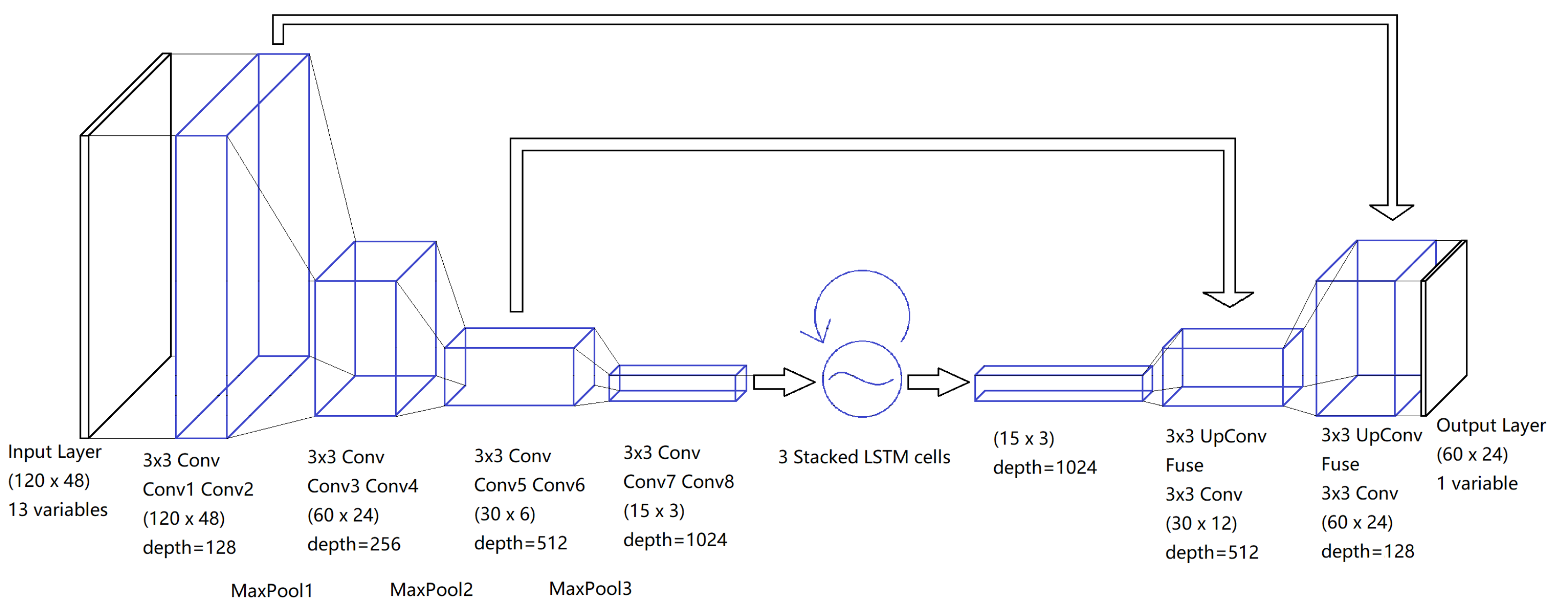}
		\caption{Deep learning model to predict JJA MDA8. The model consists of an input layer with 13 channels for the ozone predictors, 8 convolution and 3 max pooling layers to extract the dominant features in the data, and 3 stacked LSTM cells to capture the dynamics in the data. Compressed data are then passed to transposed convolution layers for projection to the output layer. The two arrows at the top indicate the skip connections that forward the high-resolution features extracted by the encoder to the decoder for better localization of the features.}
		\label{fig:model}
	\end{figure}

	
	Most of modern deep learning models are built using convolutional neural networks (CNNs) and recurrent neural networks (RNNs). CNNs are the most fundamental model in computer vision and are able to efficiently learn spatial correlations in data. The spatial correlation is captured by convolutional filtering processes in the CNN layers (described in Appendix \ref{appendix:cnn}). Another operation that is frequently applied is max pooling, which is similar to convolutional filtering, except that the convolution is replaced by a simple max transformation (described in Appendix \ref{appendix:mp}) that is used to further reduce data dimensionality and to extract dominant features. CNNs can be used as encoders that project vectors from a high-dimensional input space to a low-dimensional latent space. After each convolutional layer, the data is extracted and compressed to higher dimensional latent vectors, which are also called features or activations. A schematic of the model, which is a fully convolutional network (F-CNN), is given in Fig. \ref{fig:model}. The input layer has 13 channels for the ozone predictors. The input information gets compressed by eight convolutional blocks and three max pooling layers to extract the dominant features in the input data. 
	
	To capture the dynamics in the data we rely on the RNNs (described in Appendix \ref{appendix:rnn}). These were developed for sequential forecasting problems \cite{rnn}, and without them the dependency on the previous model state cannot be used in the later analysis. The RNN model used in this study is the long-short term memory (LSTM) cell \cite{lstm}, which is used to capture temporal correlations hidden in the data. In this study, the dynamics captured by the LSTM model includes both short-term daily variability and long-term trends in MDA8 ozone.
	The idea of combining the LSTM and CNN was first proposed by Shi et al. \cite{convlstm} for better nowcasting of precipitation. Their model showed that the hybrid deep learning model is capable of capturing short-term dynamics, which is quite challenging in the field of weather forecasting. For the prediction of summertime ozone in the United States, the problem is also challenging due to the coupling of the local meteorological conditions, the non-linear ozone chemistry, and the long-term trends in ozone. Consequently, to better capture the summertime ozone dynamics, we made the model deeper by stacking 3 LSTM cells in series. 
	
	After the input information gets compressed by the convolutional blocks, and the dynamics are captured by the LSTM cells in the final latent vector, the compressed information is projected to the output layer via a decoder that consists of a sequence of transposed convolutional layers and upsampling layers (as shown in Fig. \ref{fig:model}).  Following Ronneberger et al. \cite{unet}, we added skip connections (as show in Fig. \ref{fig:model}) that forward the high-resolution features extracted by the encoder to the decoder for better localization of the features learned by the deep learning model. High-resolution features extracted from the encoding layers are fused during the decoding process of the latent vector, which helps with the localization of important patterns \cite{unet}. Without these operations of feature fusion, the deep learning model will have difficulty in determining the most important regions where predictions should be made. The whole deep learning model has about 55 million trainable parameters, which are the convolutional kernels in the CNNs and the weights and biases in the RNNs and fully connected layers.

	The cost function to be optimized is defined as the mean squared error calculated in each grid box as follows:
	\begin{equation}
	L = \frac{1}{N} \sum_{N}^{i=1}(y_i-\hat{y}_i)^2
	\end{equation}
	where $y_i$ and $\hat{y_i}$ are the predicted and observed MDA8 ozone.
	The square of the Pearson correlation coefficient between predictions and observations is also used as an auxiliary metric of model performance,
	\begin{equation}
	r^2 = 1 - \frac{\sum_{i=1}^{N}(y_i-\hat{y}_i)^2}{\sum_{i=1}^{N}(y_i-\bar{y})^2},
	\end{equation}
	where $\bar{y}$ is the mean of observed JJA MDA8.
	This performance evaluation is only computed in grid boxes where AQS measurements are available. This way the optimization of the model is not influenced by the imperfect observational coverage of the AQS data. 
	The back-propagation algorithm is used to train the neural network (see Appendices \ref{appendix:cnn} and \ref{appendix:rnn}), with the ADAM optimization algorithm for a faster convergence \cite{adam}. 
	
	\clearpage
	\section{Evaluation of the model performance}
	\label{sec:evaluation}
	
	\begin{figure}[!h]
		\centering
		\includegraphics[width=18cm]{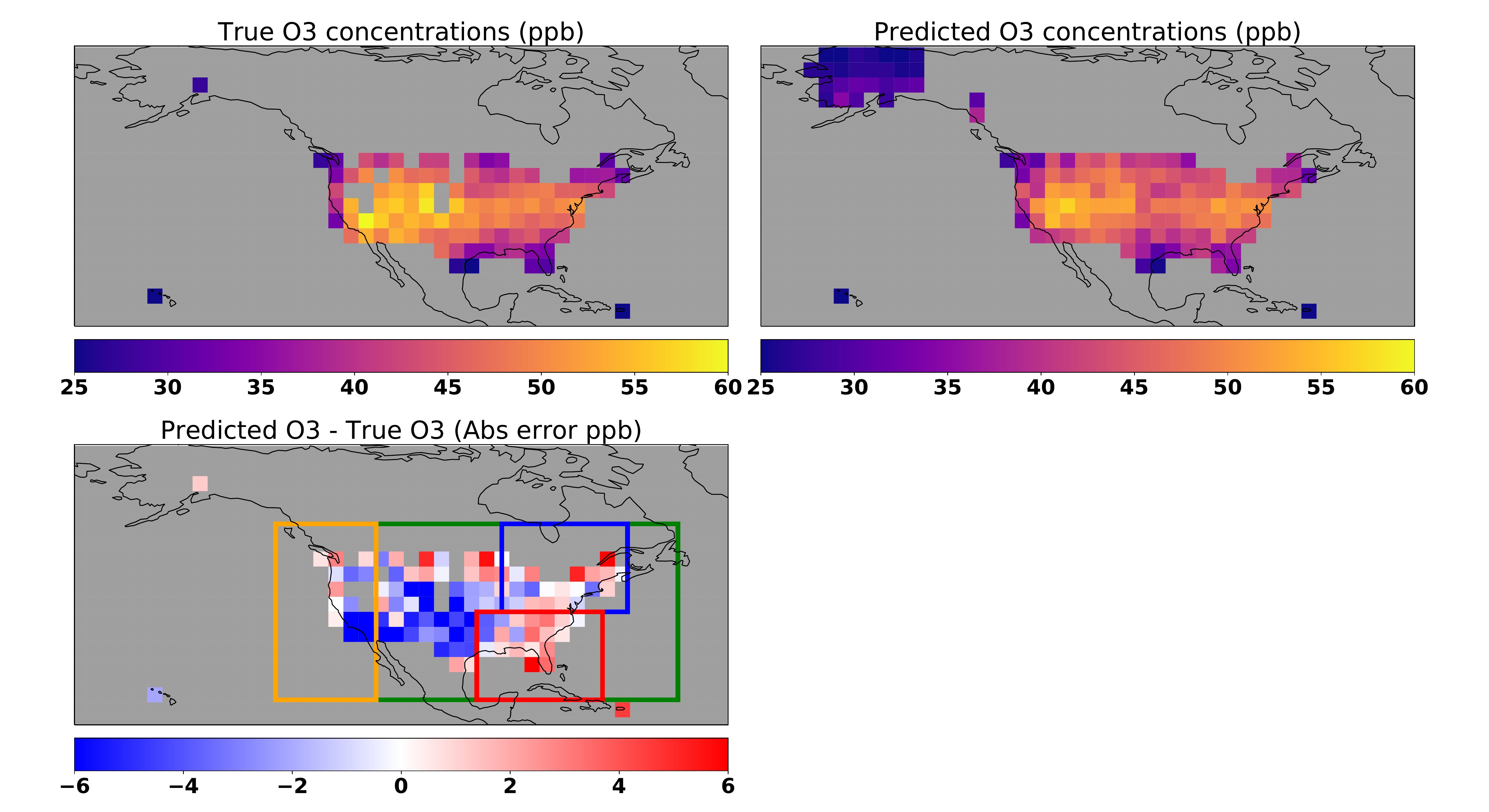}
		\caption{Observed (top left) and predicted (top right) mean JJA MDA8 ozone during 2010-2014. Also shown (bottom left) is the absolute error (in ppb) for the predicted minus observed MDA8 ozone. The boxes indicate the domains for the CONUS (green box), the northeastern (blue box) and southeastern (red box) United States, and the West Coast (yellow box) that are used in the regional analysis. Model predictions are made everywhere in the United States, but the errors are calculated only where the AQS observations are located.}
		\label{fig:meanerrs}
	\end{figure}

	\begin{figure}[!h]
		\centering
		\includegraphics[width=18cm]{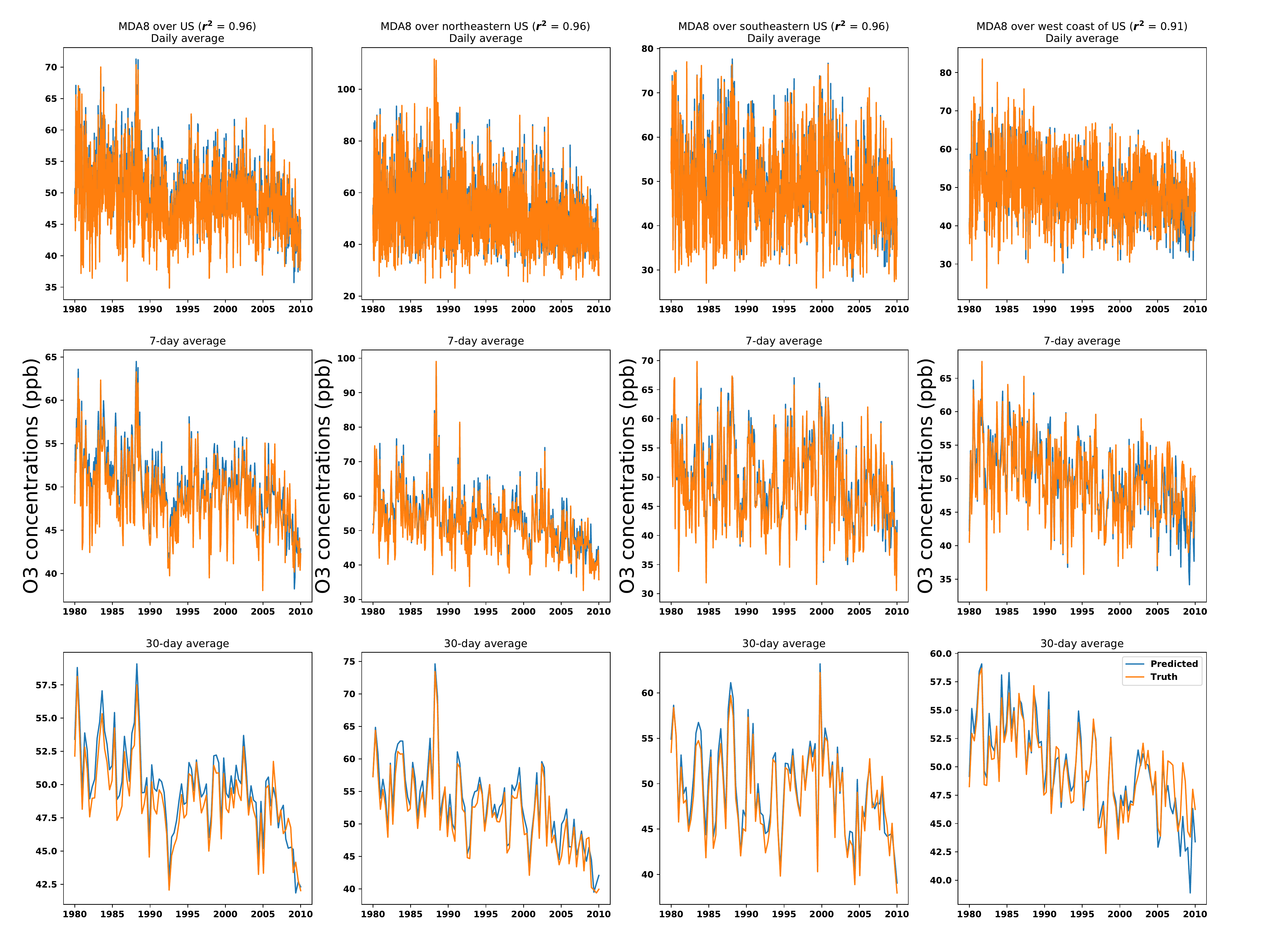}
		\caption{Observed (orange line) and predicted (blue line) daily (top row), 7-day averaged (middle row), and 30-day averaged (bottom row) JJA MDA8 ozone (in ppb) during training of the model (1980-2009). Show are the time series for the CONUS (first column), the northeast (second column), the southeast (third column), and the west coast (last column). The regional definitions are shown in Fig.~\ref{fig:meanerrs}.}
		\label{fig:traints}
	\end{figure}

	\begin{figure}[!h]
		\centering
		\includegraphics[width=18cm]{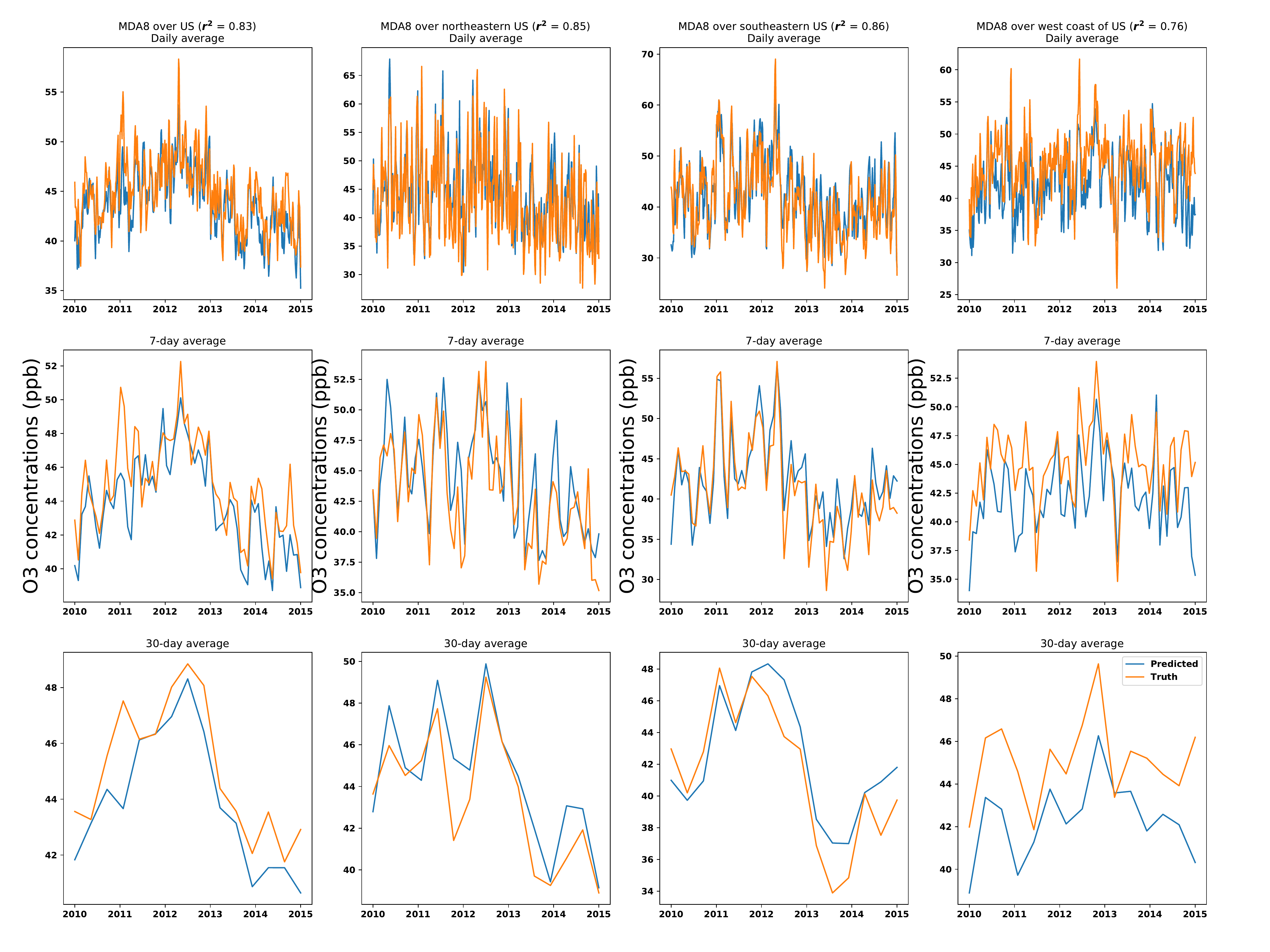}
		\caption{As in Fig.~\ref{fig:traints}, but for the testing data set (during 2010--2014).}
		\label{fig:ts}
	\end{figure}
	
	After training the model using data from 1980 to 2009, we evaluate its performance using data from the subsequent five years. As shown in Figure \ref{fig:meanerrs}, the predicted MDA8 ozone concentrations are in good agreement with the AQS ozone observations. However, the mean errors show regional differences. The largest discrepancies are in the central United States and in the Intermountain West, where the model underestimates MDA8 ozone. The model also overestimates ozone in parts of the eastern United States by 2--4 ppb.
	
	We find that the deep learning model is able to capture both the short-term and long-term dynamics of MDA8 ozone well. The time series of observed and predicted daily MDA8 ozone in the training data (for 1980 -- 2009) are plotted in Figure \ref{fig:traints} for the four regions shown in Figure \ref{fig:meanerrs} (the contiguous United States (CONUS), the northeastern and southeastern United States, and the west coast). For the CONUS, the model account for 96\% of the variability in the training data. The time series of testing data (for JJA 2010 -- 2014) are plotted in Figure \ref{fig:ts} and the error statistics are given in Table~\ref{tab:cedserrors}. The model is able to capture 85\%, 86\%, and 76\% of the variability of MDA8 in the northeastern, southeastern, and western United States, respectively. In contrast, Reidmiller et al.~\cite{reidmiller} found that atmospheric models captured only 59\%, 71\%, and 46\% of the JJA MDA8 ozone variability in the Northeast, Southeast, and California, respectively. 
	
	The mean error is less than 1 ppb in the eastern United States, which is significantly smaller than the 10--20 ppb by which conventional model simulations typically overestimate JJA MDA8 ozone in the eastern US~\cite{reidmiller}. Across the CONUS, the model underestimates MDA8 ozone by about 1 ppb. The large negative bias on the West Coast (see Table~\ref{tab:cedserrors}) and in the central United States (Fig.~\ref{fig:meanerrs}) could be due to the absence of emissions of NOx from soils in the predictors, which have been shown to be a major source of NOx in these regions \cite{jaegle, almaraz2018}. 
	
\begin{figure}[!h]
		\centering
		\includegraphics[width=12cm]{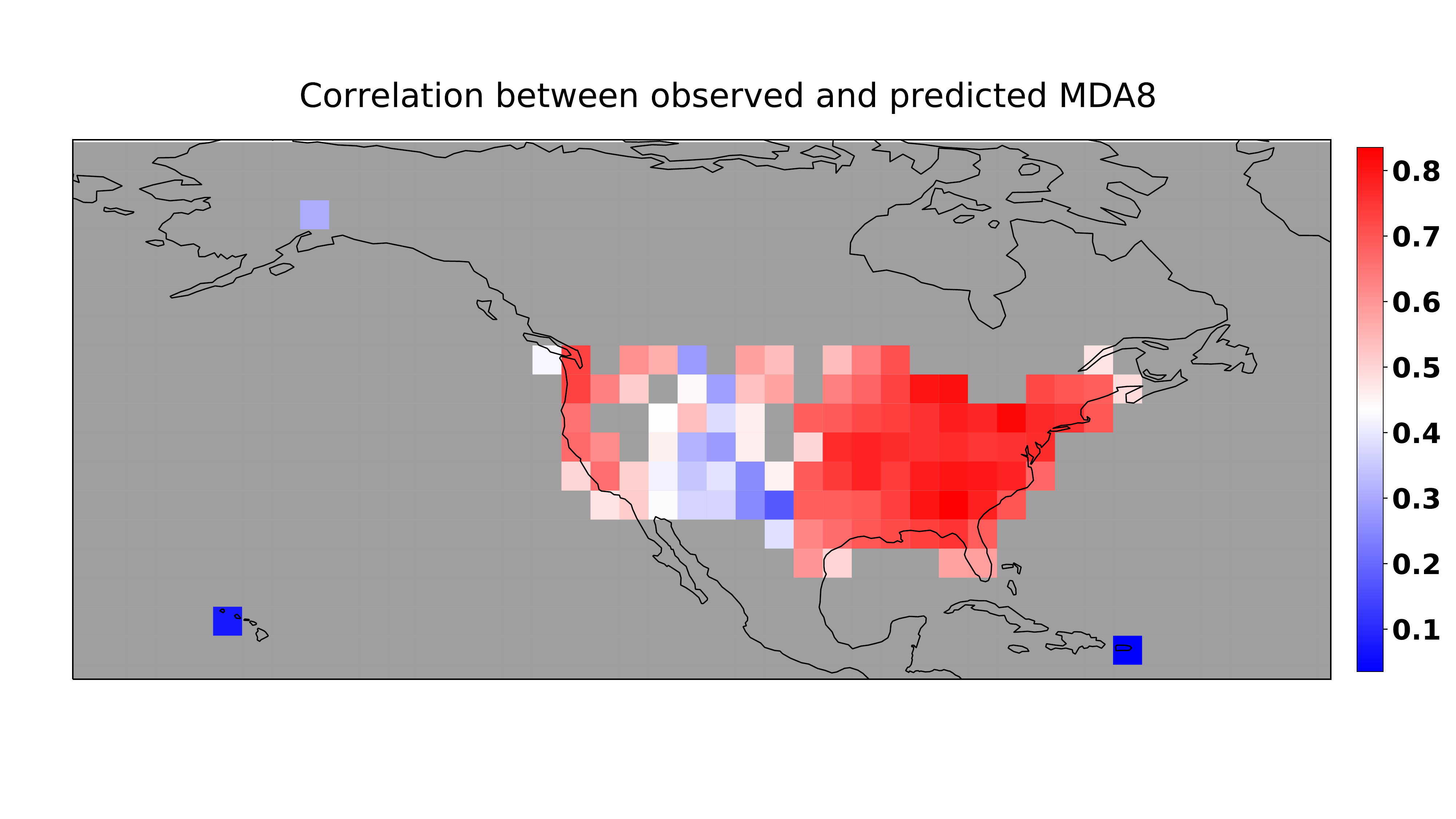}
		\caption{Correlation ($r^2$) between the observed and predicted MDA8 ozone in each grid box during the testing period (2010--2014).}
		\label{fig:corrmap}
	\end{figure}

	
	The square of the correlation coefficient for the MDA8 ozone in each grid box is plotted in Fig. \ref{fig:corrmap}. The predicted MDA8 ozone over the United States have ubiquitously high correlations ($r^2 \approx 0.75$) with the observations. However, low correlations are found in the Intermountain West ($r^2 \approx 0.4$), where there are fewer AQS observations. Also, this region is strongly influenced by free troposphere background ozone abundances rather than local or regional precursor emissions \cite{zhang2014}. Including wind fields and wildfire emissions as additional predictors may improve the predictability of MDA8 ozone in the Intermountain West, as wildfires and transport from California could have a large impact. The year-to-year variability of surface ozone is also shown to be related to stratospheric intrusions in spring \cite{zhang2014, lin-b} and the emissions of NOx from lightning in summer \cite{zhang2014}. Thus, incorporating meteorological fields related to stratospheric intrusions and lightning could potentially provide further improvement. For remote regions like Hawaii, Alaska, and Puerto Rico, shown in Fig. \ref{fig:corrmap}, the correlation is limited by the lack of sufficient observations. 
	
\begin{figure}[!h]
		\centering
		\includegraphics[width=18cm]{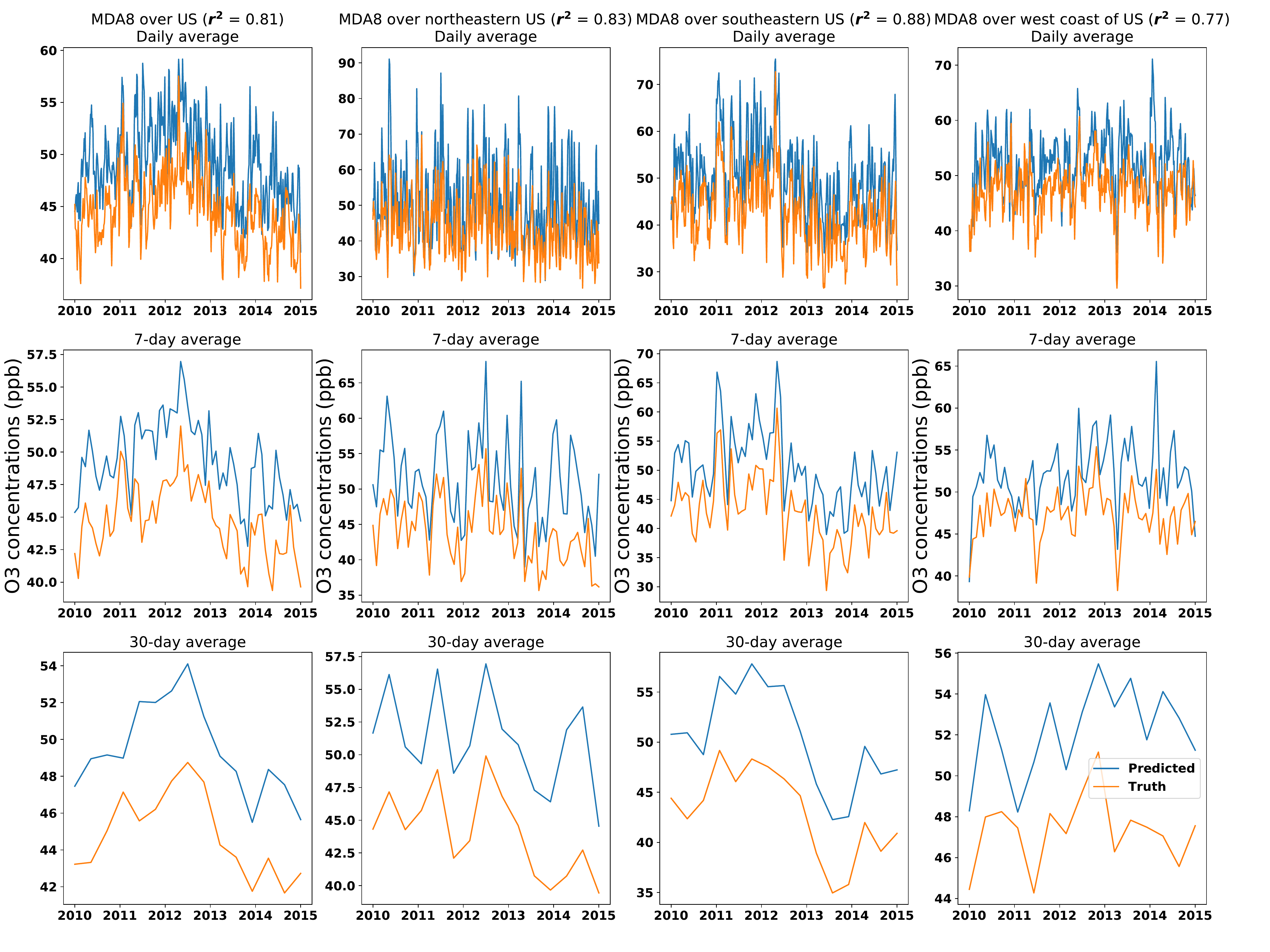}
		\caption{As in Fig.~\ref{fig:ts}, but for the experiment in which only the meteorological predictors were used in the DL model.}
		\label{fig:2010_test_ts_Met}
	\end{figure}

	We analyzed the extracted information from the fitted deep learning model (see Appendix \ref{appendix:cam_analysis}) and found that the patterns in the feature map show teleconnections between the meteorological fields, particularly over the Pacific and Atlantic oceans, and MDA8 ozone. Analysis of the NOx emission predictors indicate that the NOx emission sectors provide regional information in discriminative importance. To demonstrate the importance of the meteorological fields for capturing the ozone variability we conducted an experiment in which we trained the model with only the meteorological predictors. The results of this experiment are shown in Fig.~\ref{fig:2010_test_ts_Met} and Table \ref{tab:cedserrors}. Using only the meteorological predictors the model captures the ozone variability as well when the NOx emissions are included. For the CONUS, the model predicted MDA8 ozone with $r^2 = 0.81$ with only the meteorological predictors, compared to $r^2 = 0.83$ with the meteorological and NOx emission predictors. However, without accounting for the reductions in NOx emissions, the model overestimates ozone by about 7 ppb in the eastern US and by 4-5 ppb across the CONUS (as indicated in Table \ref{tab:cedserrors}). We note that even with this degraded performance, the fidelity of the deep learning model is still better than most atmospheric models. Including NOx emissions as a predictor is critical for capturing the long-term trend in MDA8 ozone. 
	
	The oxidation of volatile organic compounds (VOCs), in the presence of NOx, is a source of ozone that is not explicitly accounted for in the deep learning model. This is a particularly important source of ozone in the southeastern US, where there are large biogenic emissions of VOCs. But estimates of biogenic VOCs are highly uncertain. The widely used Model of Emissions of Gases and Aerosols from Nature (MEGAN), for example, overestimates biogenic VOCs in the southeastern United States \cite{millet2008}. As a result, we chose not to include biogenic VOCs as a predictor in the model. However, biogenic VOC emissions are influenced by local meteorological conditions and we expect that some of the ozone variability induced by these emissions will be captured by the meteorological predictors.
	
	In regions of high NOx emissions, ozone production will increase with emissions (biogenic and anthropogenic) of VOCs and decrease with increasing NOx emissions. This VOC-limited regime will not be captured by the model since it does not include VOC emissions as a predictor. However, we note that the main urban regions in United States, with the exception of some urban cores, are NOx-limited (ozone production increases with increasing NOx emissions) in summer, and that these regions have become more NOx-limited since 2005 as a result of reductions in NOx emissions ~\cite{duncan2010, jin2017}. Furthermore, at the coarse resolution of $3^\circ \times 3^\circ$, the deep learning model will not capture these VOC-limited urban regions. Thus, neglecting VOC emissions here should not adversely impact the performance of the model regarding changes in VOC-limited and NOx-limited regions.
	
	%
	
	\begin{table}[!h]
		\centering
		\caption{Regional error statistics for the model evaluation in the period of 2010--2014 for the model configured with the meteorological and NOx emissions predictors and for the experiment using only the meteorological predictors.}
		\begin{tabular}{l l l l l l l}
			\toprule
			&\multicolumn{2}{c}{Meteorological and NOx Predictors} & \multicolumn{2}{c}{Meteorological Predictors}\\
			\cmidrule{2-3} \cmidrule{4-5} 
			NOx trend &  Mean Error $\pm 1\sigma$ (ppb) & $r^2$ &  Mean Error $\pm 1\sigma$ (ppb) & $r^2$ \\
			\midrule
			US                               & $-1.14 \pm 1.94$  & $0.83$   & $4.59 \pm 2.24$  & $0.81$ \\
			Northeastern US         & $0.61 \pm 3.52$  & $0.85$   & $6.79 \pm 5.71$  & $0.85$\\
			Southeastern US         & $0.95 \pm 3.75$  & $0.86$   & $7.07 \pm 3.74$  & $0.88$\\
			West coast                   & $-2.66 \pm 3.23$  & $0.76$   & $4.00 \pm 3.56$  & $0.74$\\
			\bottomrule
			
		\end{tabular}
		\label{tab:cedserrors}
	\end{table}

	%

	\clearpage
	\section{Trend of anthropogenic NOx emissions over the United States after 2010}
	\label{sec:noxtrends}

\begin{figure}[!h]
		\centering
		\includegraphics[width=12cm]{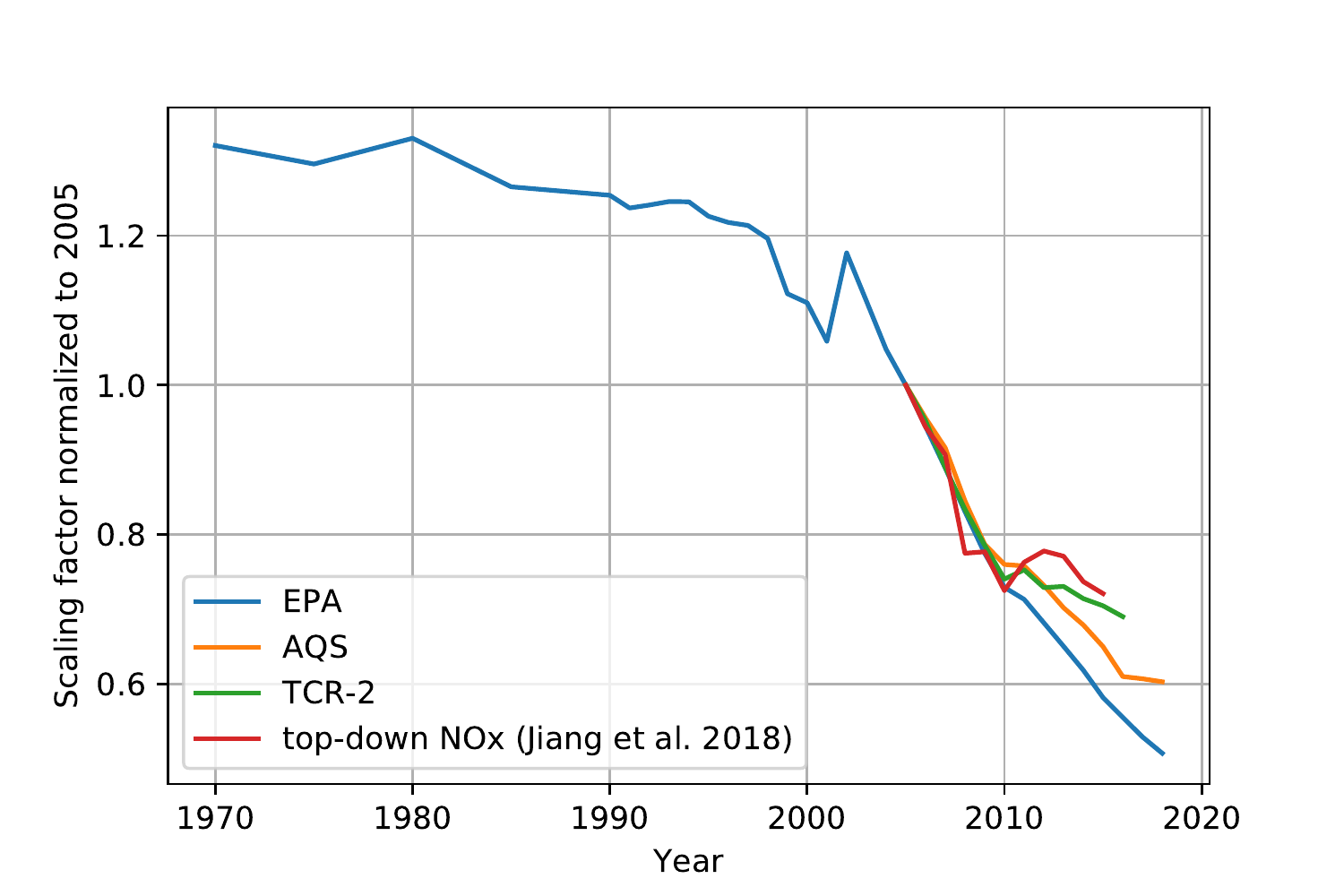}
		\caption{Relative change (normalized to 2005) in annual mean anthropogenic NOx emissions for the United States from the bottom-up EPA inventory (blue line) and from the top-down inventories from TCR-2 (green line) and Jiang et al. \cite{jiang} (red line). Also shown is the trend in AQS NO$_2$ measurements (orange line).}
		\label{fig:noxtrend}
	\end{figure}

	Figure \ref{fig:noxtrend} shows the trend in the annual mean NOx emissions from the EPA bottom-up inventory as well as from top-down emission estimates from Jiang et al. \cite{jiang} and the Tropospheric Chemistry Reanalysis (TCR-2) \cite{tcr2-a, tcr2-b, tcr2}. Compared to Jiang et al., TCR-2 used an updated data assimilation system and improved satellite retrievals. As can be seen, there is good agreement in the NOx emission trend in the different inventories between 2005, when the top-down inventories became available, and 2010. However, after 2010 the top-down inventories suggest a significant slowdown in the rate reduction of NOx emissions in the United States \cite{jiang}. Included in Fig. \ref{fig:noxtrend} is the trend in surface NO$_2$ from observations from the EPA AQS network. The AQS NO$_2$ trend suggests a smaller reduction in NOx emissions than the EPA  inventory between 2005-2010, but not as pronounced as the slowdown observed in the top-down inventories. 
	
	Evaluating these emission trends using conventional atmospheric chemical transport models is challenging due to the fact that those models are impacted by deficiencies in the employed chemical mechanisms and dynamical parameterizations. The deep learning model captures the physical and chemical mechanisms between MDA8 and its predictors based on the input \textit{in situ} and meteorological data only, and is able to mitigate the impact of a majority of sources of error in conventional atmospheric models. Here we use the deep learning model to assess the consistency of the bottom-up and top-down NOx emission inventories with observed MDA8 ozone.
	
	To evaluate the trends in the NOx emissions, we use the trained model to predict MDA8 ozone from 2010 to 2016 using the CEDS NOx emissions scaled by the different annual trends shown in Fig. \ref{fig:noxtrend}. The CEDS inventory is scaled as follows:
	\begin{equation}
	E^{m}_{i} = E^{CEDS}_{i}*\beta^{m}
	\end{equation}
	where $E^{CEDS}_{i}$ is the CEDS emissions in month $i$, $\beta^{m}$ is the annual scaling factor that captures the trend shown in Fig. \ref{fig:noxtrend} for a given inventory $m$, and $E^{m}_{i}$ is the resulting scaled NOx emissions used in the model prediction of MDA8 ozone. 
	
\begin{figure}[!h]
		\centering
		\includegraphics[width=18cm]{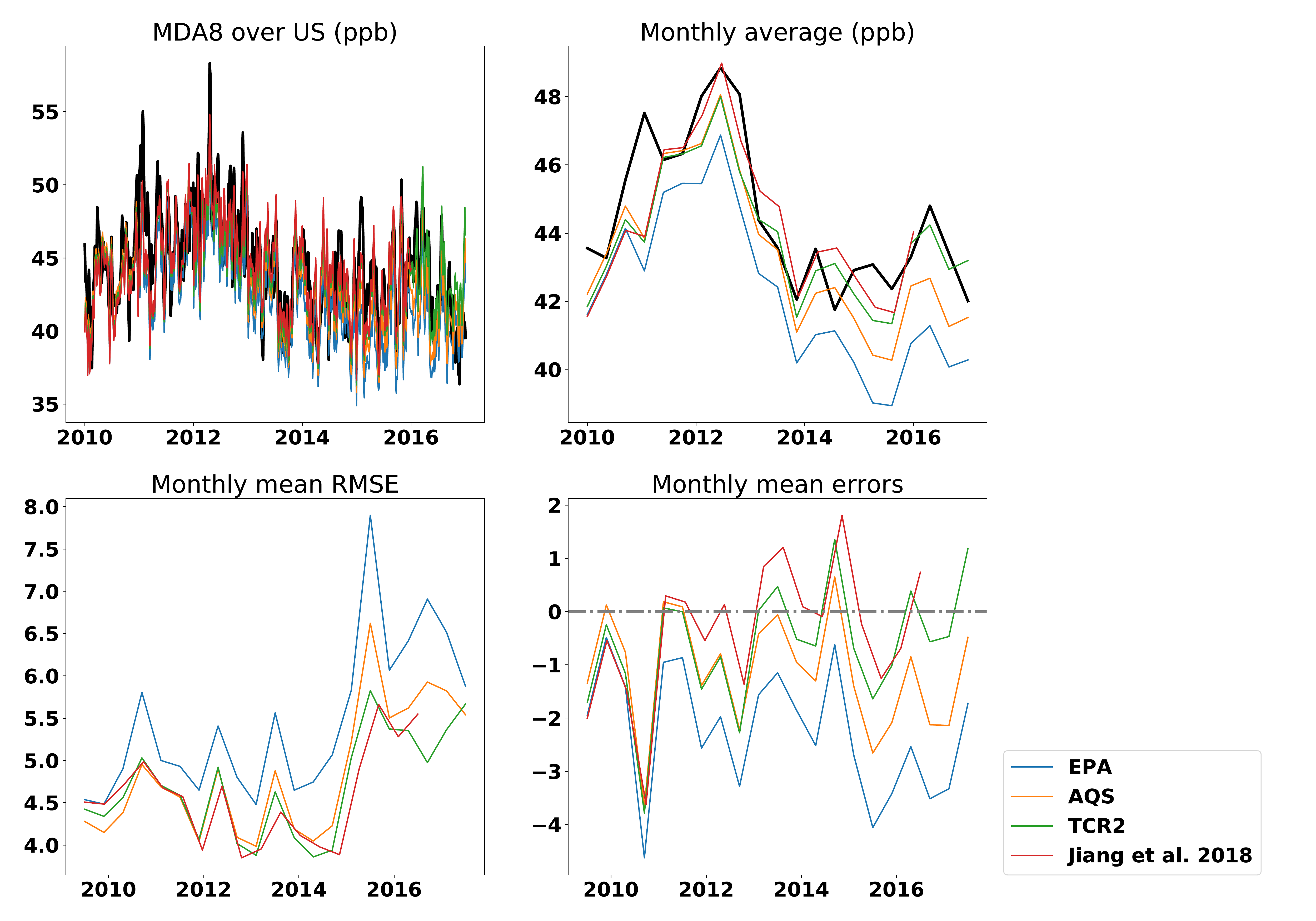}
		\caption{Top row: Observed and predicted daily mean (top left) and monthly mean (top right) MDA8 ozone between 2010--2016 (2010--2015 for Jiang et al.). Shown are the AQS ozone observations (black line) and the model predictions based on the NOx emissions scaled by the EPA (blue line), AQS NO$_2$ (orange line), TCR-2 (green line), and the Jiang et al. (red line) trends. Bottom row: The monthly RMSE (bottom left) and monthly mean errors (bottom right) for the MDA8 ozone predictions in the first row.}
		\label{fig:noxanalysis}
	\end{figure}

	The time series of predicted and observed MDA8 ozone are plotted in Fig. \ref{fig:noxanalysis} and the error statistics are shown in Table \ref{tab:errortable}. The observed AQS NO$_2$ trend results in a mean error of $-1.06 \pm 2.04$ ppb across the United States, which is statistically indistinguishable from the standard results obtained with the CEDS inventory (in Fig. \ref{fig:ts}). In contrast, the EPA trend results in a larger bias of $-2.18 \pm 2.10$ ppb, which is clearly visible in Fig. \ref{fig:noxanalysis}. The EPA trend also results in the largest root-mean-square errors (RMSE) (see Fig. \ref{fig:noxanalysis}). The TCR-2 trend produces the smallest error relative to the MDA8 ozone observations.
	
	The greater consistency of the top-down trends with the surface ozone observations could be due to the fact that inversion analyses that produced these emission estimates incorporated satellite observations of tropospheric ozone and NO$_2$, so the inferred NOx emissions reflect the trends in both tropospheric ozone and NO$_2$. We note that in a sensitivity analysis (see Appendix \ref{appendix:control_exp}) in which we retrained the model using data from 1980-2005 and predicted MDA8 ozone for 2005-2016, the predicted ozone based on the bottom-up and top-down trends were consistent with each other and with the observed AQS ozone between 2005-2010. After 2010, however, the EPA trend produced the largest negative bias in predicted ozone, whereas the top-down trends were in better agreement with observed AQS ozone observations. 
	
	It was suggested \cite{silvern, li_2019_trends} that the discrepancy between the top-down and bottom-up NOx emission estimates could be due to the fact that the satellite observations of NO$_2$ are more representative of non-anthropogenic NOx in rural regions after 2010. In Table~\ref{tab:urbanstats} the error statistics for the model predictions for 2010--2016 are aggregated in "urban" and "rural" regions, defined according to whether NOx emissions in a given grid box are greater than or less than $1 \times 10^{11}$ molec cm$^{-2}$ s$^{-1}$, respectively, following Li and Wang \cite{li_2019_trends}. The NOx emissions scaled by the observed AQS NO$_2$ trend produce ozone predictions with the smallest errors and highest correlations in the urban regions. For the rural regions, the best performance is obtained with the top-down NOx trends, which show smaller differences between urban and rural regions. Our results agree with previous studies \cite{silvern, li_2019_trends}, suggesting that the top-down NOx trends are affected by a combination of anthropogenic NOx emissions and rural NOx conditions. Our results also show that the bottom-up EPA trend is more inconsistent with observed ozone than either the AQS or top-down trend. 
	
	\begin{table}[!h]
		\centering
		\caption{MDA8 ozone error statistics for the CONUS for 2010--2016 (2010--2015 for Jiang et al.).}
		\begin{tabular}{l l l l l}
			\toprule
			NOx trend &  Mean Error $\pm 1\sigma$ (ppb) & $r^2$ \\
			\midrule
			EPA     & $-2.18 \pm 2.10$  & $0.80$ \\
			AQS            & $-1.06 \pm 2.04$  & $0.81$ \\
			TCR-2         & $0.55 \pm 2.10$  & $0.79$ \\
			Jiang et al.   & $-0.69 \pm 1.21$  & $0.80$ \\
			\bottomrule
			
		\end{tabular}
		\label{tab:errortable}	
	\end{table}

	\begin{table}[!h]
		\centering
		\caption{Urban and rural MDA8 ozone error statistics for the CONUS for 2010--2016 (2010--2015 for Jiang et al.).}
		\begin{tabular}{l l l l l l l}
			\toprule
			&\multicolumn{2}{c}{Urban} & \multicolumn{2}{c}{Rural}\\
			\cmidrule{2-3} \cmidrule{4-5} 
			NOx trend &  Mean Error $\pm 1\sigma$ (ppb) & $r^2$ &  Mean Error $\pm 1\sigma$ (ppb) & $r^2$ \\
			\midrule
			EPA         & $-1.36 \pm 3.41$  & $0.81$   & $-2.36 \pm 2.16$  & $0.76$ \\
			AQS         & $-0.08 \pm 3.43$  & $0.82$   & $-1.28 \pm 2.09$  & $0.78$\\
			TCR-2       & $0.47 \pm 3.62$  & $0.79$   & $-0.79 \pm 2.10$  & $0.76$\\
			Jiang et al.& $0.67 \pm 3.67$  & $0.80$   & $-0.57 \pm 2.04$  & $0.78$\\
			\bottomrule
			
		\end{tabular}
		\label{tab:urbanstats}
	\end{table}
	

	\clearpage
	\section{Conclusions}
	\label{sec:conclusions}
	We have developed an ozone prediction system based on state-of-the-art deep learning models to predict summertime daily MDA8 ozone in the United States. The model uses 13 predictors, including large-scale meteorological variables and sector-specific anthropogenic emissions of NOx. The model was trained with observed summertime MDA8 ozone data from 1980 to 2009 and tested with data from 2010 to 2014. We found that the model captured well the daily variability in MDA8 ozone across the United States, predicting ozone with $r^2 = 0.83$ and a mean error of $-1.14$ ppb. Regionally, the model has high predictability of ozone in the eastern United States and on the west coast ($r^2 \approx 0.75$), but low predictability in the Intermountain West ($r^2 \approx 0.4$). The negative bias in largest in the west coast (with a mean error $-2.66$ ppb), whereas the model has a slight positive bias (less than $+1$ ppb) in the eastern United States. Feature maps show that the deep learning model captures the teleconnections between MDA8 ozone and the meteorological predictors, which are responsible for driving the daily, seasonal, and interannual variations in predicted ozone. The NOx emissions are important for capturing the long-term negative trend in surface ozone in the United States.
	
	We used the model to evaluate recent trends in NOx emissions after 2010 in the context of the model predictions of surface ozone. We found that between 2010 to 2016 the trends in NOx emissions in the bottom-up EPA inventory resulted in the largest underestimate of MDA8 ozone across the United States (with a mean error $-2.18$ ppb). In contrast, the trend in NOx emissions consistent with the trend in AQS NO$_2$ observations resulted in summertime MDA8 ozone predictions that were in better agreement with the AQS ozone observations (with a mean error of $-1.06$ ppb), and were consistent with those obtained with the CEDS bottom-up inventory. The best agreement with the AQS ozone observations were obtained with the top-down NOx emission trends from TCR-2 \cite{tcr2-a, tcr2-b, tcr2} and Jiang et al. \cite{jiang}. Examination of the error statistics aggregated into urban and rural regions revealed that in urban regions the AQS trend provided ozone predictions in agreement with observations, whereas in rural regions the top-down trends produced the best agreement with observations. Our results suggest that the top-down trends are capturing changes in anthropogenic and non-anthropogenic NOx after 2010. The EPA trend produced the largest errors in ozone in both urban and rural regions, suggesting that the EPA inventory is overestimating the reduction in NOx emissions after 2010.
	
	This deep learning architecture is generic. It can be utilized to realize other high-dimensional predictions, given the spatial and temporal dynamics in the data. Although we aggregated the MDA8 ozone data to a resolution of $3^\circ \times 3^\circ$,  the model is able to deal with various spatial resolutions. Despite the greater predictive skill of our deep learning approach compared to conventional atmospheric chemical transport models, other modern machine learning algorithms have the potential for even greater gains in performance. For example, deep Gaussian process (GP) models are frequently used in modelling temporal dynamics. To mitigate the impact of limited observational coverage, generative adversarial networks (GANs) could be used to interpolate data gaps using the learned correlations between predictions and predictors. These algorithms could offer superior performance for air quality and other Earth Science applications.
	
	\section{Acknowledgments}
This work was supported by the Natural Sciences and Engineering Research Council of Canada. Computations were performed on the Graham supercomputer of Compute Ontario and Compute Canada. Part of this work was conducted at the Jet Propulsion Laboratory, California Institute of Technology, under contract with the National Aeronautics and Space Administration (NASA).
	
	\clearpage
	\begin{appendices}
		
		\section{Convolutional neural networks and error back-propagation}
		\label{appendix:cnn}
		For a set of two-dimensional kernel matrices K, the convolutional filtering process could be expressed as:
		\begin{eqnarray} 
		O_{i,j,k} &=& (A*K_k)_{i,j} = \sum_{m}\sum_{n} [A(i-m, j-n)K_k(m, n)] \\
		S_{i, j, k} &=& G( {O}_k )_{i, j}
		\end{eqnarray}
		where the asterisk $*$ denotes the convolution operation, and $m$ and $n$ are the indices along the two major axes of the input data $A$. The result of the convolution calculation $O_{i,j,k}$ is transformed by the activation function $G$. $i$ and $j$ denote the indexes along the major axes of the output tensor $S$, which is referred to as the feature map. $k$ is the kernel index.
		
		The optimization of CNNs is done using the back-propagation algorithm \cite{dnn, lecun}. To back propagate error information through CNNs, the partial derivatives of cost function with respect to the convolutional and pooling kernels are:
		\begin{eqnarray} 
		(\frac{\partial J}{\partial K_k})_{m, n} &=& ( \sum_{i}\sum_{j} \frac{\partial J}{\partial S_{i,j,k}} \frac{\partial S_{i,j,k}}{\partial O_{i,j,k}} \frac{\partial O_{i,j,k}}{\partial K_k})_{m,n} \nonumber \\
		&=& ( \sum_{i}\sum_{j} (\hat{S}_{i,j,k} - S_{i,j,k}) G'( {O}_k )_{i, j}  A(i-m, j-n) )_{m,n} \nonumber \\
		&=& (\delta S_k \cdot G'( {O}_k )* A)_{m,n}  
		\end{eqnarray}
		where $\hat{S}_k$ is the true output, and $\delta S_k$ denotes the entrywise differences between prediction and output. Here chain rule and the commutativity of convolution operation are applied. At each learning step, the kernels could be optimized along the gradients as:
		\begin{eqnarray} 
		K_k'(m, n) = K_k(m, n) - \eta (\frac{\partial J}{\partial K_k})_{m, n} \nonumber
		\end{eqnarray}
		where $\eta$ is the size of the learning rate, which could be optimized by a sensitivity test.
		We can see that the optimization of a convolutional layer could be expressed as another convolution operation.
		
		\section{Max pooling layers}
		\label{appendix:mp}
		The max pooling process could be expressed as:
		\begin{eqnarray} 
		S_{i, j, k} = max\{ A_{m,n,k} : i \le m < i+p_1, j \le n < j + p_2 \}
		\end{eqnarray}
		where $p_1$ and $p_2$ stand for the dimension of the max filter.
		
		\section{Recurrent neural networks and error back-propagation}
		\label{appendix:rnn}
		The flow of information in the LSTM cell is filtered by three gates: forget, input, and output gates. At each time step, these gates estimate prior errors based on the previous model state, and controls the amount of information from new observations to be stored in the cell state.
		
		The feed-forward process of the LSTM network at time $t$ could be expressed as:
		\begin{eqnarray} 
		f_t &=& \sigma(w_f [h_{t-1}, x_t] + b_f)  \\
		i_t &=& \sigma(w_i [h_{t-1}, x_t] + b_i) \\
		o_t &=& \sigma(w_o [h_{t-1}, x_t] + b_o) \\
		\tilde{C}_t &=& \tanh{(w_C [h_{t-1}, x_t] + b_C)} \\
		{C}_t &=& f_t C_{t-1} + i_t \tilde{C}_t \label{eq:3.73} \\
		h_t &=& o_t \tanh{(C_t)} 
		\end{eqnarray}
		where $x_t$ and $h_t$ stand for the input and output vectors at time $t$. $\tilde{C}_t$ and ${C}_t$ denote the prior and posterior model states, which were designed to specifically characterize the latent state of dynamics.
		
		The first three equations represent the computation made at forget, input and output gates, respectively. The input vector at time $t$, $x_t$, is concatenated with output vector from previous analysis $h_{t-1}$ to form an augmented state, which is then transformed by the corresponding weight and bias term at each gate. These gates are normalized by the sigmoid activation function $\sigma$ to be between 0 and 1, where 0 means the gates are blocking everything and vice versa. The forget rate $f_t$ and input rate $i_t$ could be thought of as the estimated errors associate with previous posterior model state ${C}_{t-1}$ and the prior model state at current time step $\tilde{C}_t$. And the new posterior model state ${C}_{t}$ is then computed by solving Equation (\ref{eq:3.73}). The final posterior analysis $h_t$ is computed by activating the posterior model state, and is weighted by the output rate $o_t$. Here the all the indexes are omitted, and all the operation is entrywise.
		
		To embed the LSTM cells with other neural networks, the back propagation of error gradients has to be formulated. The training algorithm used for LSTM units is called back propagation through time (BPTT):
		
		\begin{eqnarray} 
		\frac{\partial J}{\partial w_f} &=& \frac{\partial J}{\partial h_t} \frac{\partial h_t}{\partial C_t} \frac{\partial C_t}{\partial f_t} \frac{\partial f_t}{\partial w_f} \nonumber \\
		&=& \delta h_t o_t sech^2(C_t) C_{t-1} [h_{t-1}, x_t] \sigma'(w_f [h_{t-1}, x_t] + b_f) \nonumber \\
		\frac{\partial J}{\partial b_f} &=& \frac{\partial J}{\partial h_t} \frac{\partial h_t}{\partial C_t} \frac{\partial C_t}{\partial f_t} \frac{\partial f_t}{\partial b_f} \nonumber \\
		&=& \delta h_t o_t sech^2(C_t) C_{t-1} \sigma'(w_f [h_{t-1}, x_t] + b_f) \nonumber  \\
		\frac{\partial J}{\partial w_i} &=& \frac{\partial J}{\partial h_t} \frac{\partial h_t}{\partial C_t} \frac{\partial C_t}{\partial i_t} \frac{\partial i_t}{\partial w_i} \nonumber \\
		&=& \delta h_t o_t sech^2(C_t) \tilde{C}_t [h_{t-1}, x_t] \sigma'(w_i [h_{t-1}, x_t] + b_i) \nonumber \\
		\frac{\partial J}{\partial b_i} &=& \frac{\partial J}{\partial h_t} \frac{\partial h_t}{\partial C_t} \frac{\partial C_t}{\partial i_t} \frac{\partial i_t}{\partial b_i} \nonumber \\
		&=& \delta h_t o_t sech^2(C_t) \tilde{C}_t \sigma'(w_i [h_{t-1}, x_t] + b_i) \nonumber \\
		\frac{\partial J}{\partial w_o} &=& \frac{\partial J}{\partial h_t} \frac{\partial h_t}{\partial o_t} \frac{\partial o_t}{\partial w_o} \nonumber \\
		&=& \delta h_t \tanh(C_t) [h_{t-1}, x_t] \sigma'(w_o [h_{t-1}, x_t] + b_o) \nonumber  
		\end{eqnarray}
		\begin{eqnarray} 
		\frac{\partial J}{\partial b_o} &=& \frac{\partial J}{\partial h_t} \frac{\partial h_t}{\partial o_t} \frac{\partial o_t}{\partial b_o} \nonumber \\
		&=& \delta h_t \tanh(C_t) \sigma'(w_o [h_{t-1}, x_t] + b_o) \nonumber  \\
		\frac{\partial J}{\partial w_C} &=& \frac{\partial J}{\partial h_t} \frac{\partial h_t}{\partial C_t} \frac{\partial C_t}{\partial \tilde{C}_t} \frac{\partial \tilde{C}_t}{\partial w_C} \nonumber \\
		&=& \delta h_t o_t sech^2(C_t) i_t [h_{t-1}, x_t] sech^2(w_C [h_{t-1}, x_t] + b_C) \nonumber   \\
		\frac{\partial J}{\partial b_C} &=& \frac{\partial J}{\partial h_t} \frac{\partial h_t}{\partial C_t} \frac{\partial C_t}{\partial \tilde{C}_t} \frac{\partial \tilde{C}_t}{\partial b_C} \nonumber \\
		&=& \delta h_t o_t sech^2(C_t) i_t sech^2(w_C [h_{t-1}, x_t] + b_C) \nonumber   
		\end{eqnarray}
		where $\delta h_t$ denotes the output error at time $t$. The training starts from the cost function calculated at last time step and propagates backwards. The optimization process will run iteratively along the whole data sequence, until the cost function is fully minimized.
		
		\clearpage
		\section{Analysis on teleconnections between ozone predictability and large-scale circulation patterns}
		\label{appendix:cam_analysis}
		
\begin{figure}[!h]
			\centering
			\includegraphics[width=12cm]{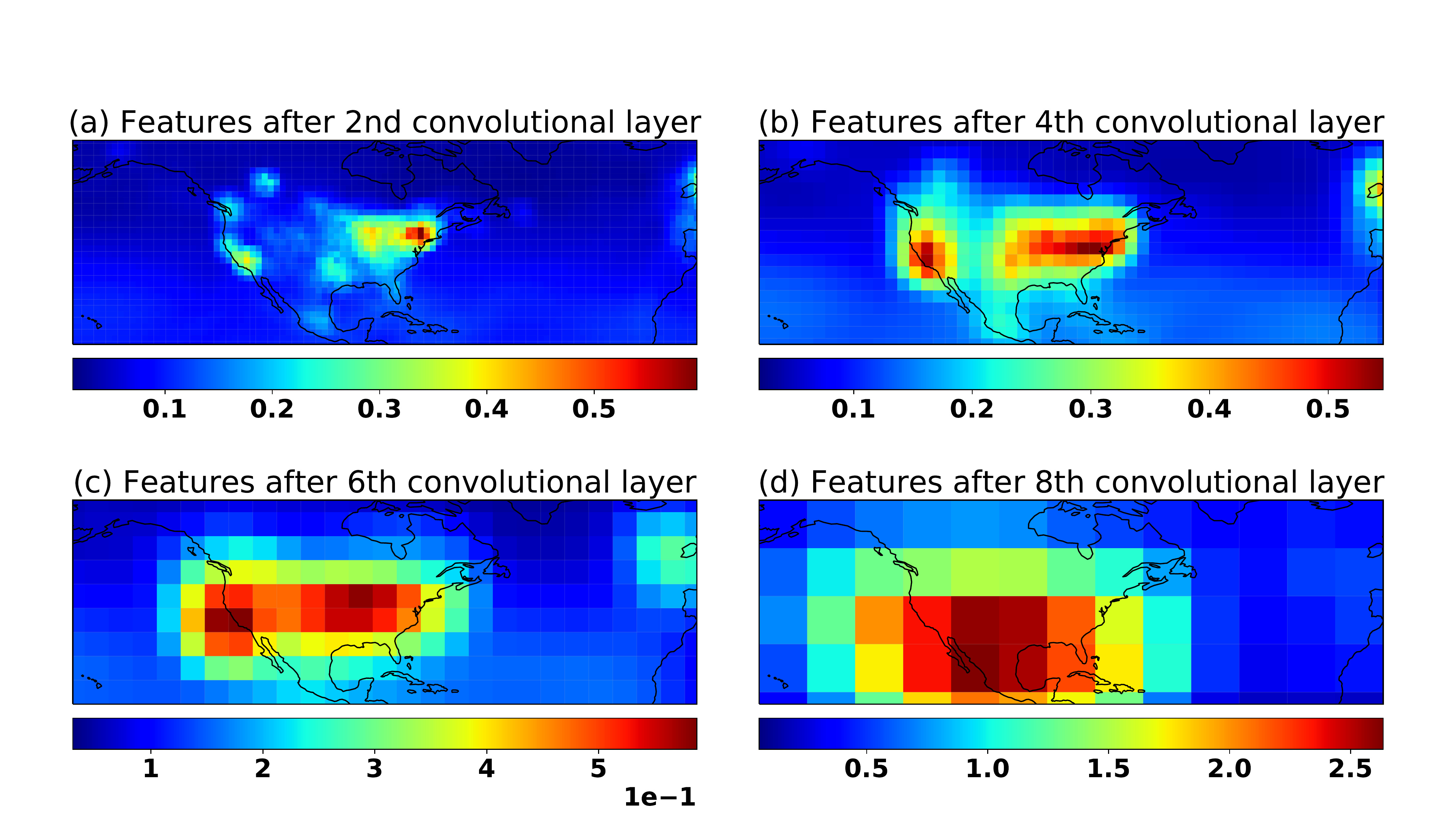}
			\caption{Feature maps from the 2nd, 4th, 6th, and 8th convolutional layers as illustrated in Fig. \ref{sec:model}, averaged within all training data from 1980 to 2009.}
			\label{fig:actmap}
		\end{figure}

		The convolutional kernels in each layer in the deep learning model control the amount of information to be passed into subsequent layers, which are optimized along the gradients of the cost function. Although the features extracted by the convolutional layers are unitless and difficult to directly relate to physical parameters, they could be used as high-dimensional representation of hidden correlations with the model output. We plot in Fig. \ref{fig:actmap} the the feature maps from the Conv2, Conv4, Conv6, and Conv8 convolution layers from the encoder part of the deep learning model illustrated in Fig. \ref{fig:model}. Throughout the process of information compression, consistent strong features are located over the continent, corresponding to the local influence of NOx emissions. In Fig. 8 (a) and (b), some activated features could also be found in the northeastern Pacific Ocean, which represents teleconnections between large-scale circulation patterns and MDA8 ozone \cite{lshen, sutton1, sutton2, gill}. 
		

\begin{figure}[!h]
			\centering
			\includegraphics[width=11cm]{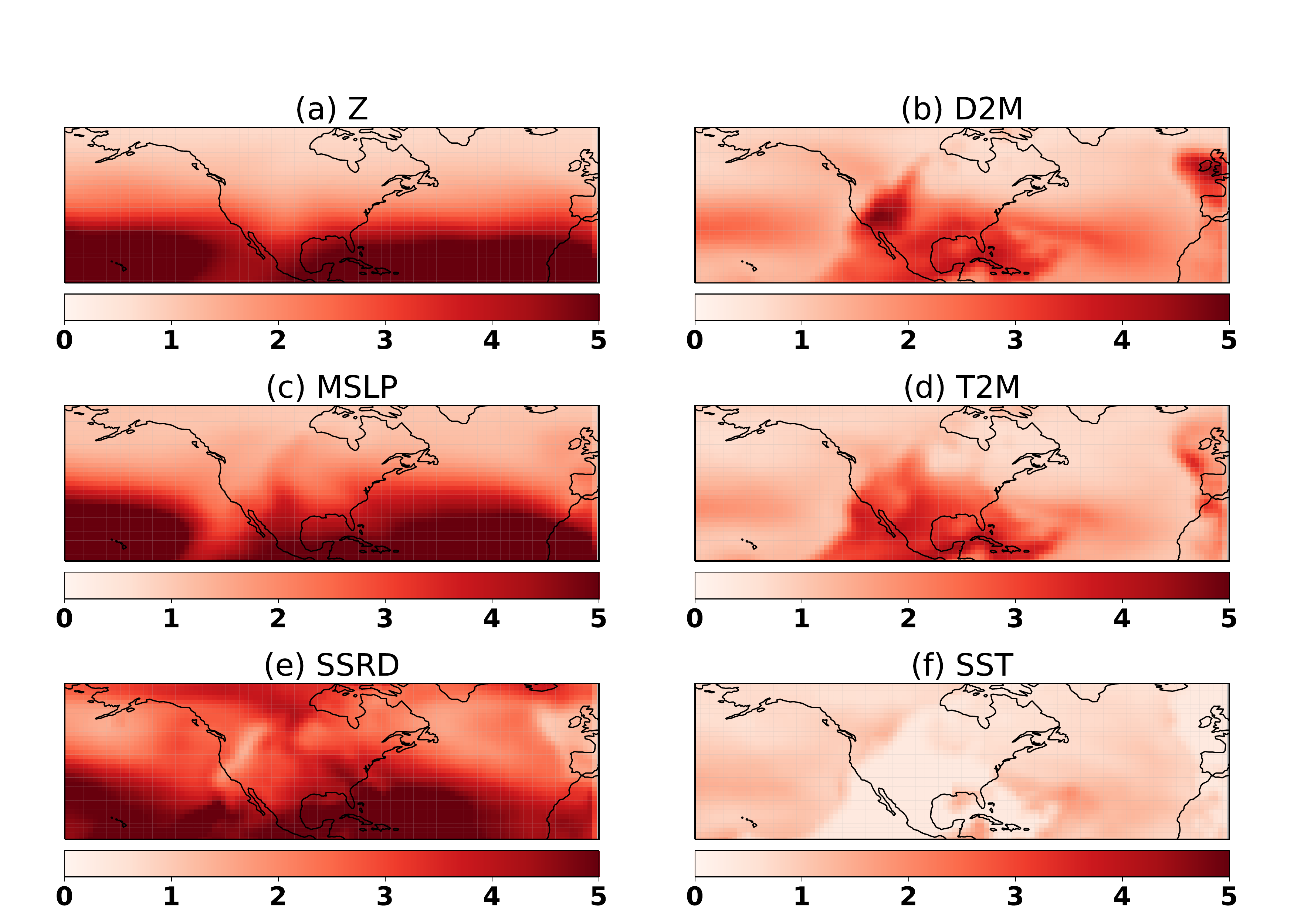}
			\caption{Averaged feature maps for the meteorological predictors, calculated from the training data set.}
			\label{fig:metact}
		\end{figure}

		Feature maps from the first convolutional layer are of particular interest because they represent the sensitivity of the predicted JJA MDA8 ozone, and could be distinguished between each input predictors. The features can be utilized to analyze the discriminative sensitivity of the deep learning model to each ozone predictor. Inspired by \cite{cnnfeatures, cam}, we compute the channel-wise activation maps (ChAM) to analyze the significance of each ozone predictor in ozone predictability, which is defined as
		
		\begin{equation}
		\alpha^{c}_{i, j} = \frac{1}{N} \sum_{k = 1}^{N}( w^{c}_k \star x^{c})_{i, j}
		\end{equation},
		
		where $c$ is the index of ozone predictors, $k$ is the index of convolutional kernels, and $\star$ denotes the convolutional operation. We compute ChAM using the testing data set and plot the average feature maps for the meteorological predictors in Fig. \ref{fig:metact}. We find that for Z, MSLP, SSRD and SST, the central Pacific Ocean and the Atlantic Ocean are the most discriminative regions for ozone predictability. For D2M and T2M, the largest features are mainly located over the southwest and the Gulf Coast. Remarkably, even if SST is only defined over the ocean, the convolutional operation has a smearing effect and the model can still capture the continental teleconnections between MDA8 and SST. All meteorological predictors show sensitivity around Hawaii and in the Atlantic, which coincides with previous study on teleconnections between MDA8 with SST and MSLP\cite{lshen}.
		
		
\begin{figure}[!h]
			\centering
			\includegraphics[width=12cm]{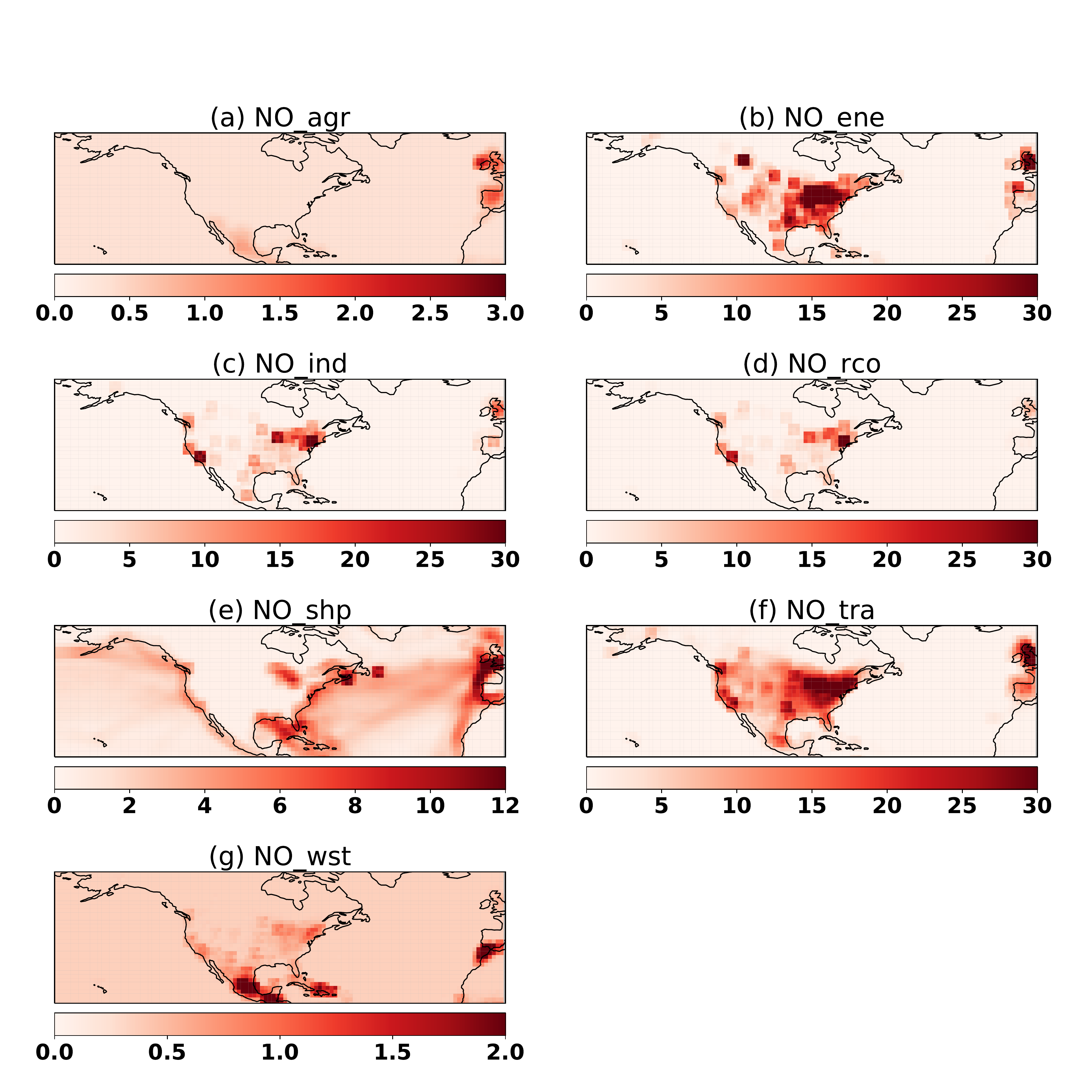}
			\caption{Average activation maps for NOx emissions, calculated from training data set. The seven NOx emission sectors are as follows: (a) agriculture (AGR), (b) the power industry (ENE), (c) manufacturing (IND), (d) residential and commercial (RCO), (e) international shipping (SHP), (f) surface transportation (TRA), and (g) waste disposal (WST).}
			\label{fig:noxact}
		\end{figure}

		Unlike meteorological predictors, the activations of NOx emissions are very local, which is likely due to the short lifetime of NOx near the surface. Fig. \ref{fig:noxact} shows the ChAM for the NOx emission sectors. The NOx emissions related to agriculture and waste handling are mainly affecting ozone predictability in the southern US, although on average they have less activation in the deep learning model compared to other sectors. Overall, ENE, IND and TRA are the most important contributors to ozone predictability.
		
		\clearpage
		\section{Restricting training period from 1980-2009 to 1980-2005}
		\label{appendix:control_exp}

\begin{figure}[!h]
			\centering
			\includegraphics[width=18cm]{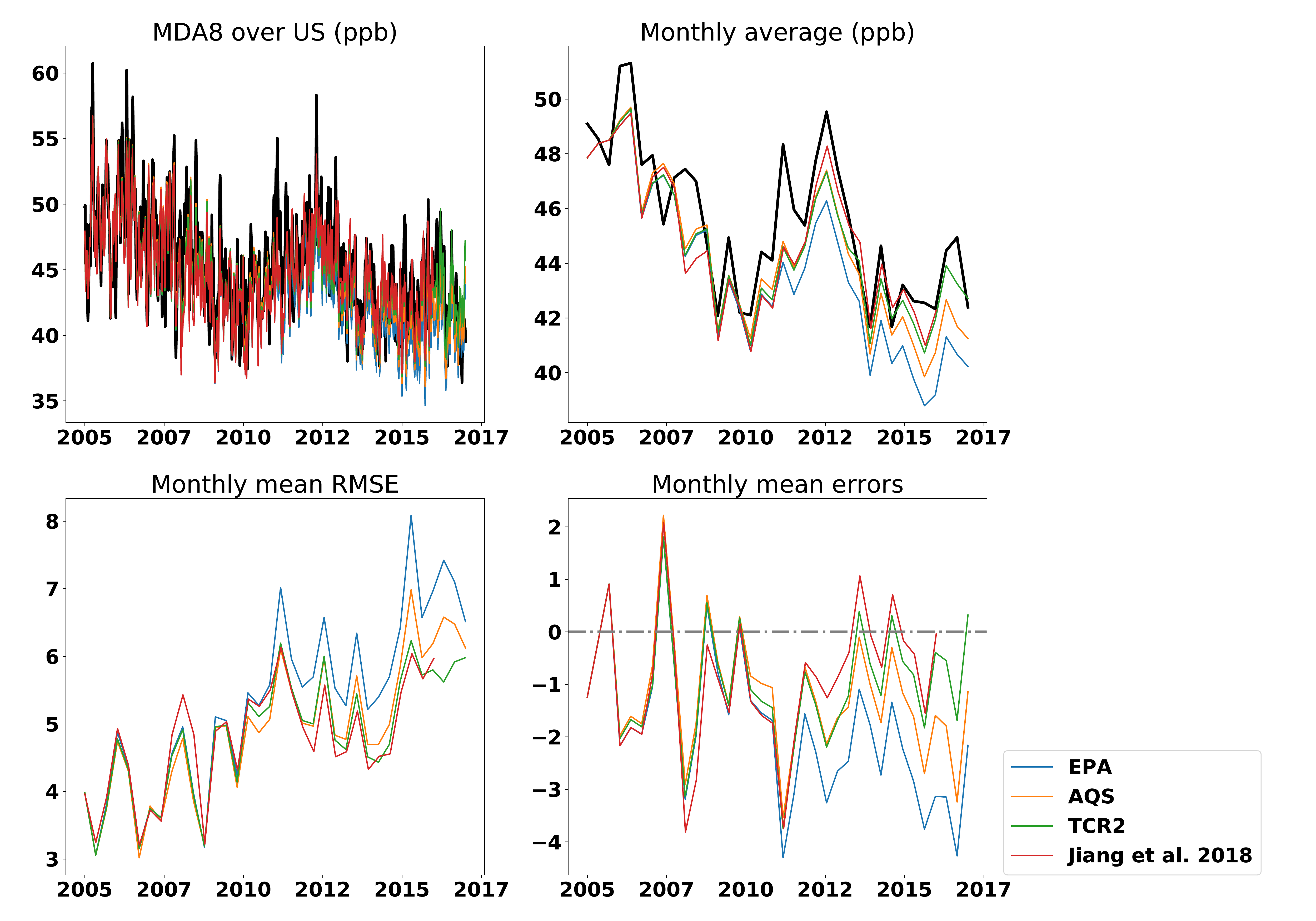}
			\caption{Top row: Observed and predicted daily mean (top left) and monthly mean (top right) MDA8 ozone between 2005--2015 (2005--2015 for Jiang et al.). Shown are the AQS ozone observations (black line) and the model predictions based on the NOx emissions scaled by the EPA (blue line), AQS NO$_2$ (orange line), TCR-2 (green line), and the Jiang et al. (red line) trends. Bottom row: The monthly RMSE (bottom left) and monthly mean errors (bottom right) for the MDA8 ozone predictions in the first row.}
			\label{fig:noxanalysis_exp}	
		\end{figure}

		To evaluate the model for the 2005-2016, we retrained the model from 1980 to 2005, with the same other experimental settings as in Section \ref{sec:noxtrends}, and tested it from 2005-2016. The time series of the predicted and observed MDA8 are plotted in Fig. \ref{fig:noxanalysis_exp}, and the error statistics for 2005-2009 and 2010-2016 are given in Tables \ref{tab:prior2010error} and \ref{tab:post2010error}, respectively. Between 2005-2010, the MDA8 ozone predicted using the different NOx trends all show good consistency over the US. However, after 2010, the bottom-up trends of NOx resulted in an underestimation of MDA8 ozone relative to tat from the top-down trends. The divergence is clearly visibly in the time series of the monthly mean errors in Fig. \ref{fig:noxanalysis_exp}, with the EPA-based trend clearly producing the largest RMSE and negative bias after 2010. The degraded performance for the 2010-2016 period relative to the results in Section \ref{sec:noxtrends} is due to the reduced length of the training period.
		
		

		\clearpage
		
		\begin{table}[!h]
			\centering
			\caption{MDA8 ozone error statistics for the CONUS for 2005--2009.}
			\begin{tabular}{l l l l l}
				\toprule
				NOx trend &  Mean Error $\pm 1\sigma$ (ppb) & $r^2$ \\
				\midrule
				EPA             & $0.10 \pm 2.75$  & $0.78$ \\
				AQS            & $0.32 \pm 2.76$  & $0.78$ \\
				TCR-2         & $0.18 \pm 2.73$  & $0.78$ \\
				Jiang et al.   & $0.00 \pm 2.82$  & $0.78$ \\
				\bottomrule
				
			\end{tabular}
			\label{tab:prior2010error}	
		\end{table}
		
		\begin{table}[!h]
			\centering
			\caption{MDA8 ozone error statistics for the CONUS for 2010--2016 (2010--2015 for Jiang et al.).}
			\begin{tabular}{l l l l l}
				\toprule
				NOx trend &  Mean Error $\pm 1\sigma$ (ppb) & $r^2$ \\
				\midrule
				EPA             & $-1.83 \pm 2.32$  & $0.74$ \\
				AQS            & $-0.84 \pm 2.28$  & $0.75$ \\
				TCR-2         & $-0.46 \pm 2.39$  & $0.72$ \\
				Jiang et al.   & $-0.29 \pm 2.34$  & $0.74$ \\
				\bottomrule
				
			\end{tabular}
			\label{tab:post2010error}	
		\end{table}

	\end{appendices}

	\clearpage
	\bibliographystyle{unsrt}  
	
	\newpage

\end{document}